\documentclass[conference]{IEEEtran}

\usepackage{cite}
\usepackage[cmex10]{amsmath}
\usepackage{xcolor}
\usepackage{algorithmic}
\usepackage{array}  
\usepackage{mdwmath}
\usepackage{mdwtab}
\usepackage{eqparbox}
\usepackage[tight,footnotesize]{subfigure}
\usepackage{booktabs}
\usepackage{url}
\usepackage[colorlinks]{hyperref}
\newcommand{\cut}[1]{}

\newcommand{\eg}{\textit{e}.\textit{g}.}

\ifCLASSINFOpdf
  \usepackage[pdftex]{graphicx}
  \graphicspath{{./imgs/}}
\else
  \usepackage[dvips]{graphicx}
\fi

\begin{document}
%
\title{Security of Deep Learning based Lane Keeping System under Physical-World 
Adversarial Attack\vspace{-0.2in}}

\author{\IEEEauthorblockN{
Takami Sato\thanks{\IEEEauthorrefmark{1}The first two authors contributed equally.}\IEEEauthorrefmark{1},
Junjie Shen\IEEEauthorrefmark{1},
Ningfei Wang, 
Yunhan Jack Jia\IEEEauthorrefmark{2},
Xue Lin\IEEEauthorrefmark{3}, and
Qi Alfred Chen}
\IEEEauthorblockA{Univesity of California, Irvine; \IEEEauthorrefmark{2}ByteDance AI Lab;
\IEEEauthorrefmark{3}Northeastern University
}
}

\IEEEoverridecommandlockouts

\maketitle

\begin{abstract}
Lane-Keeping Assistance System (LKAS) is convenient and widely available today, but also extremely security and safety critical. In this work, we design and implement the first systematic approach to attack real-world DNN-based LKASes. We identify dirty road patches as a novel and domain-specific threat model for practicality and stealthiness. We formulate the attack as an optimization problem, and address the challenge from the inter-dependencies among attacks on consecutive camera frames. We evaluate our approach on a state-of-the-art LKAS and our preliminary results show that our attack can successfully cause it to drive off lane boundaries within as short as 1.3 seconds.

%

\end{abstract}

\section{Introduction}
\label{sec: intro}




Lane-Keeping Assistance System (LKAS) is an Level-2 driving automation technology that automatically steers a vehicle to keep it within the current traffic lane~\cite{sae2018}
. Due to its high convenience for human drivers, today it is widely available in a variety of vehicle models such as Honda Civic, Toyota Prius, Nissan Cima, Volvo XC90, Mercedes-Benz C-class, Audi A4, and Tesla Model S.
For models without such function yet, there are also solutions such as OpenPilot and Ghost that can retrofit them to support LKAS after adding cheap hardware such as a front camera~\cite{openpilot}.
While convenient, such function is extremely security and safety critical: When LKAS starts to make wrong steering decisions, an average driver reaction time of 2.3 seconds~\cite{driver_reaction_time2000}
may not be enough to prevent the vehicle from colliding into vehicles in adjacent lanes or in opposite directions, or driving off road to hit road curbs or fall down the highway cliff. Even with collision avoidance systems, it cannot prevent the vehicle from hitting the curb, falling down the highway cliff, or being hit by other vehicles that fail to yield. Thus, it is urgent and highly necessary to understand the security property of LKAS.

To achieve lane keeping, the most critical step in an LKAS is lane detection, which by default uses camera due to the nature of lane lines. So far, Deep Neural Network (DNN) based detection achieve the state-of-the-art accuracy ~\cite{tusimple}
and is adopted in the most performant LKASes today such as Tesla Autopilot and OpenPilot~\cite{openpilot}. Thus, the end-to-end security of the latest LKAS technology highly depends on the security of such DNN models. While recent works show that DNN models are vulnerable to carefully crafted input perturbations~\cite{Szegedy2014, goodfellow2014explaining}
, their methods cannot be directly applied to attack DNN-based LKASes due to 3 unique challenges. First, prior methods are mostly designed for classification or object detection, and none of their attack formulations can be directly applied for lane detection. Second, to affect the camera input of an LKAS, the perturbations need to be realizable in the physical world and can normally appear on traffic lane regions. Moreover, such perturbations must not affect the original human-perceived lane information for stealthiness. Prior works have explored such threats for traffic signs~\cite{eykholt2018robust
, zhao2018seeing}, but not for traffic lanes. 


Third, to cause end-to-end impact to an LKAS, the attack needs to affect a sufficient number of consecutive camera frames, and most importantly, the attacks on later frames are dependent on those on earlier frames. For example, if the attack successfully deviates the detected lane to the right in a frame, the LKAS will control the vehicle heading accordingly, which causes the following frames to capture road areas more to the right and thus directly affect their attack generation. To the best of our knowledge, no prior work considers attacking a sequence of image frames with such strong inter-dependencies.


The only prior effort that successfully attacked an LKAS is from Tencent~\cite{tencent2019}, where they fooled the Tesla DNN-based LKAS to follow fake lane lines created by a line of big white dots on road regions originally without lane lines. However, it is neither attacking the scenarios where an LKAS is designed for, i.e., roads with lane lines, nor generating the perturbations systematically by addressing all the three challenges above.

To fill this critical research gap, in this work we design and implement the first systematic approach to attack real-world DNN-based LKASes. To practically introduce perturbations, we identify road patches as the threat model, which is specific to lane detection models and can normally appear in the physical world. For stealthiness, we restrict the perturbations to be within lane lines, and the color space to be on the gray scale to pretend to be a benign but dirty road patch. We then formulate the malicious road patch generation as an optimization problem, and design a multi-frame path bending objective function specifically for the lane detection task. To address the challenge from the inter-dependencies among attacks on consecutive camera frames, we design a novel car motion model based input generation process and a gradient aggregation technique.

We evaluate our approach on a state-of-the-art LKAS, OpenPilot, and our preliminary results show that our attack can successfully deviate an LKAS to drive off the lane boundaries within as short as 1.3 seconds, which is far shorter than 2.3 seconds, the average driver reaction time~\cite{driver_reaction_time2000}. To better illustrate the attack effect, we also prepare a short demo from the victim car driver's view at our project website \textbf{\url{https://sites.google.com/view/lane-keeping-adv-attack/}}.


\section{Threat Model and Problem Formulation}

\textbf{Threat model}. We assume that the attacker can possess the same LKAS as the one in the victim vehicles and has the full knowledge of the LKAS via reverse engineering. Before attacking, the attacker can also collect camera frames on the target road by driving her own vehicle with the LKAS. 

\begin{figure}[!t]
\centering
\includegraphics[width=3.4in]{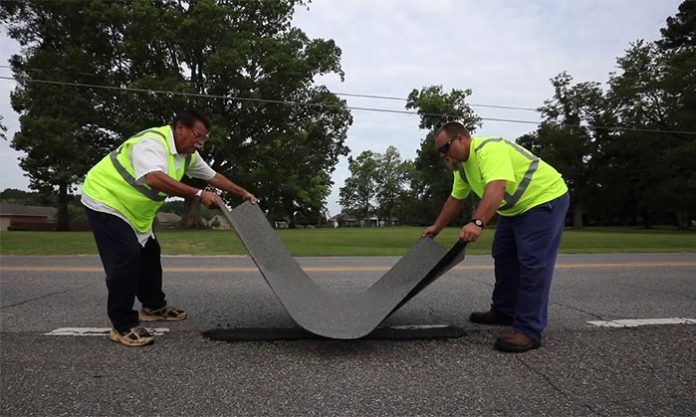}
\caption{An example of our novel and domain-specific threat model: dirty road patches. The attacker can add malicious input perturbations on top of such adhesive road patches~\cite{adhesive_patch}.}
\label{fig:threat_model}
\end{figure}

\textbf{Dirty road patch: realizable and stealthy physical-world attack vector.} We identify malicious road patches as the attack vector, since they are realizable in the physical world and can normally appear around traffic lanes. For stealthiness, we restrict these road patches to not cover the original lane lines and their color to be on the gray scale to pretend to be benign but dirty. The attacker can print the malicious input perturbations on asphalt, rubber, or poster, and then place it on the road. Fig.~\ref{fig:threat_model} shows one possible real-world attack scenario that places an adhesive road patch to introduce malicious input perturbations to LKAS. Such adhesive road patches are widely available in the US~\cite{adhesive_patch}.

\textbf{Attack goal and safety damages}. By attacking the LKAS, we aim to cause the victim car to have a lateral deviation large enough to drive out of the current lane boundaries within the common driver reaction time, which thus fundamentally breaks the design goal of LKAS and can cause severe safety consequences.
Assuming the victim vehicle locates at the lane center before the attack, the required deviation is 0.745 meters on the highway in the US~\cite{road_width} and the average driver reaction time is 2.3 seconds~\cite{driver_reaction_time2000}.

The attack goal above can directly cause various types of safety hazards in the real world:

\vspace{-\topsep}
\begin{itemize}
\setlength{\itemsep}{0pt}
\setlength{\parskip}{0pt}
\item \textit{Driving off road}, which is a direct violation of traffic rules~\cite{offroad_violate_law} and can cause various safety hazards such as hitting road curbs or falling down the highway cliff. These cannot be prevented even when the vehicle can perform perfect obstacle and collision avoidance. 

\item \textit{Vehicle collisions}, \eg, with vehicles parked on the road side, or driving in adjacent or opposite traffic lanes. Even when the vehicle can perform obstacle and collision avoidance, these collisions are still possible for two reasons. First, today's obstacle and collision avoidance systems are not perfect. For example, a recent study shows that automatic braking systems in popular car models today fail to avoid crashes 60\% of the time~\cite{aeb-fail}. Second, even if they can successfully perform emergency stop, they cannot prevent the victim vehicle from being hit by other vehicles that fail to yield on time, given that human drivers have an 2.3 seconds average reaction time~\cite{driver_reaction_time2000}.

\vspace{-\topsep}
\end{itemize}

\textbf{Attack incentives.} No matter whether road accidents are actually caused in the end, causing the victim cars to exhibit unsafe driving behaviors or violate traffic rules can already damage the reputation of the corresponding LKAS system providers, e.g., the corresponding car manufacturers. Thus, a likely attack incentive is \textit{business competition}, which can allow one LKAS system provider to deliberately damage the reputation of its rival companies and thus unfairly gain competitive advantages. Meanwhile, considering the direct safety impact, we also cannot rule out the possible incentives for terrorist attacks, \eg, for political or financial purposes.




\begin{figure}[!t]
\centering
\includegraphics[width=3.4in]{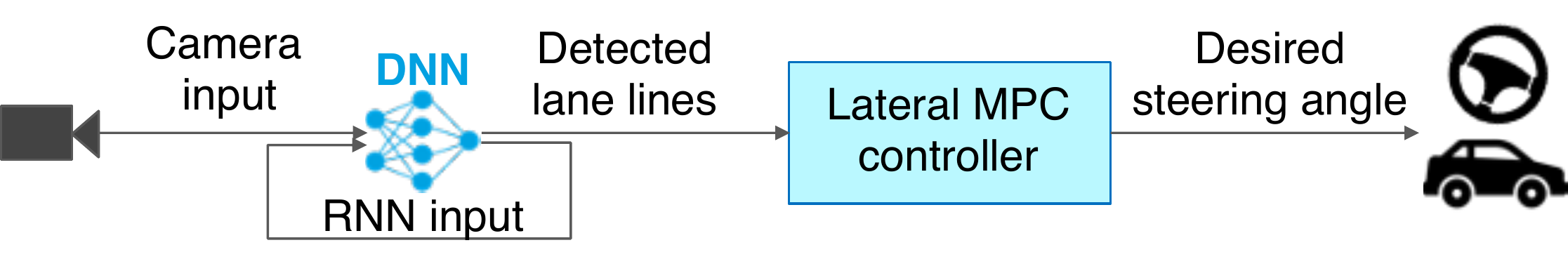}
\caption{OpenPilot LKAS pipeline.}
\label{fig:openpilot_LKAS}
\end{figure}


\section{Attack Methodology}
\label{sec:attack_method}

With the problem formulation above, we design the following novel techniques to address the challenges in~\S\ref{sec: intro}.

\begin{figure}[!t]
\centering
\includegraphics[width=3.4in]{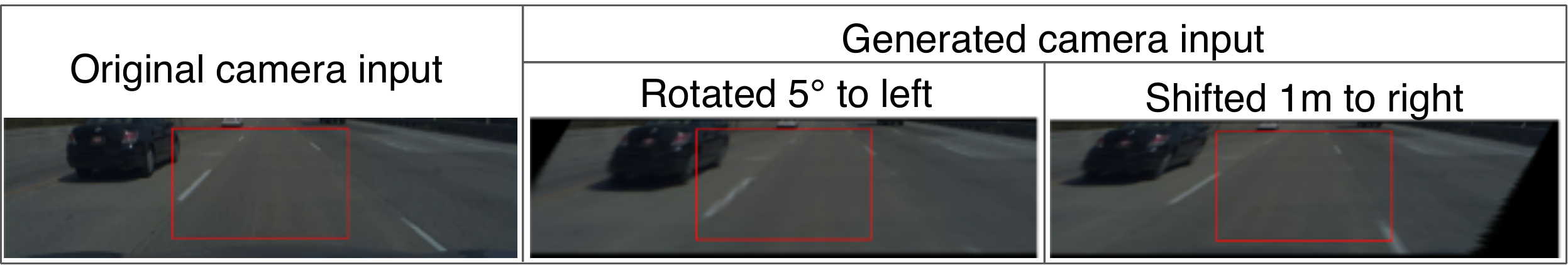}
\caption{Car motion model based camera input generation from the original camera input. The red rectangle denotes the model input area.}
\label{fig:input_generation}
\end{figure}

\textbf{Car motion model based input generation}. To consider the inter-dependencies among attacks on consecutive camera frames, we need to dynamically update camera inputs according to the driving trajectory changes during the patch generation.
To address this, we use a bicycle model~\cite{bicyclemodel}
to simulate the changes to car trajectory, which is then used to update camera inputs by applying perspective transformations to the original non-attacked camera inputs.
Fig.~\ref{fig:input_generation} shows an example of this generation process. On the bird's eye view (BEV), we apply a car position shift and heading angle change from the original car trajectory and then project the BEV image back to the camera perspective. Although it causes some distortion and partial missing area, the model input area, which locates at the center, is still complete and usable.

\textbf{Multi-frame path bending objective function}. To generate the malicious road patch, we adopt an optimization-based method, which has shown both high efficiency and effectiveness in previous works. Since the lateral controller of the LKAS is not differentiable, we introduce a surrogate objective function to deviate the car as much as possible. 
The lateral controller calculates a \textit{desired driving path} based on the detected lane lines, and numerically solves a steering angle plan to enforce this path. The desired driving path is typically represented by a polynomial function. Assuming the car strictly follows the path, the derivatives of the path are essentially the wheel angles it needs to apply. Thus, we formulate our objective function:

\begin{figure} [h]
\scriptsize
\begin{align}
\scriptsize
    f(X_1,...,X_T, s_0) =
    \sum_{t=1}^{T}\sum_{d \in D} \nabla p_{t}(d; { \{X_j| j \leq t\}, s_0}) + \lambda ||\Omega_t(X_t)||^2_2
    \label{math:objective}
\end{align}
\end{figure}

, where $p_t(x)$ is the desired driving path in the $t$-th frame, $X_t$ is the $t$-th generated camera inputs including the malicious road patch, $s_0$ is the initial state and $D$ is the set of the points where the controller makes steering angle decisions, $\lambda$ is a weight of the L2 regularization, and $\Omega$ is a function that extracts the patch area in a camera input. $p_t(x)$ is decided by the image inputs of current and previous camera inputs $ \{X_j| j \leq t\}$ and the initial state $s_0$ through the DNN model. We minimize this objective function when attacking to the right and maximize it when attacking to the left.

\begin{figure}[!t]
\centering
\includegraphics[width=3.4in]{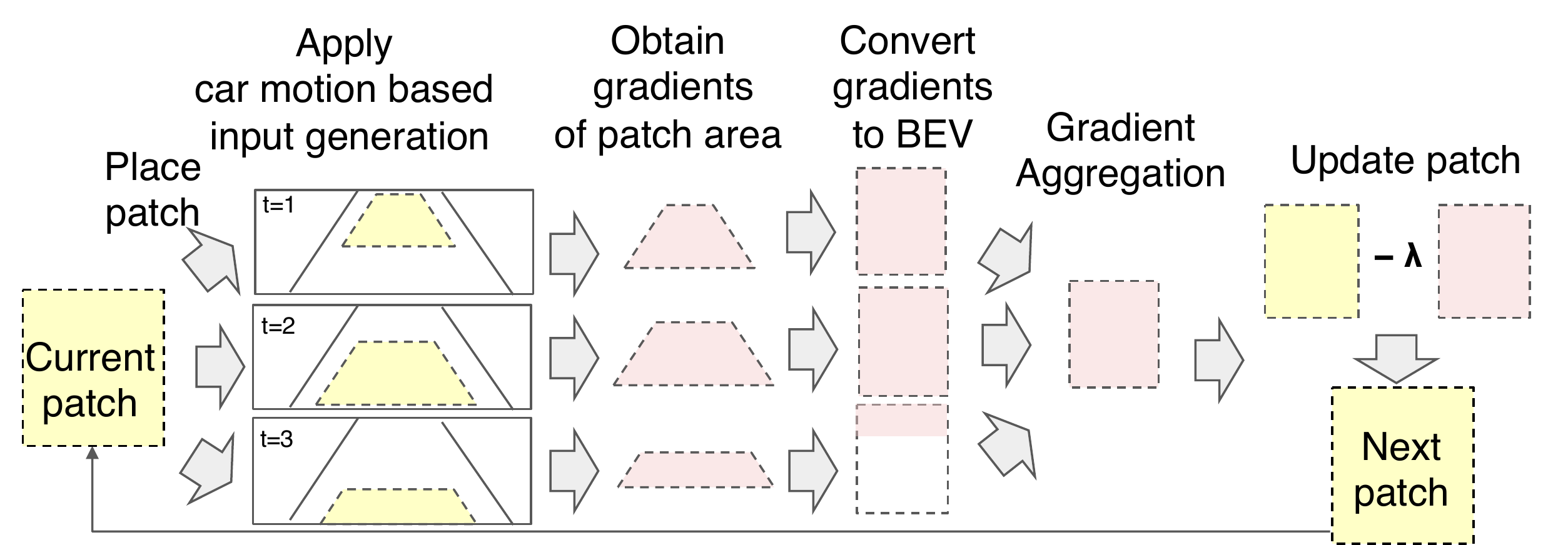}
\caption{Overview of the optimization pipeline of the malicious road patch.}
\label{fig:grad_agg}
\end{figure}

\textbf{Gradient aggregation}. Based on the objective function, we obtain the gradients of each camera input.
However,  the gradient descent is not directly applicable to update the malicious road patch since the patch sizes and portions are different in each camera input. To address this, we transform all camera inputs to BEV to align gradients to the same scale and take a weighted average as shown in Fig.~\ref{fig:grad_agg}. In addition,  we restrict update directions to the gray scale in order to pretend to be a benign but dirty road patch for stealthiness.

\section{Early Results}

\begin{figure}[!t]
\centering
\includegraphics[width=3.4in]{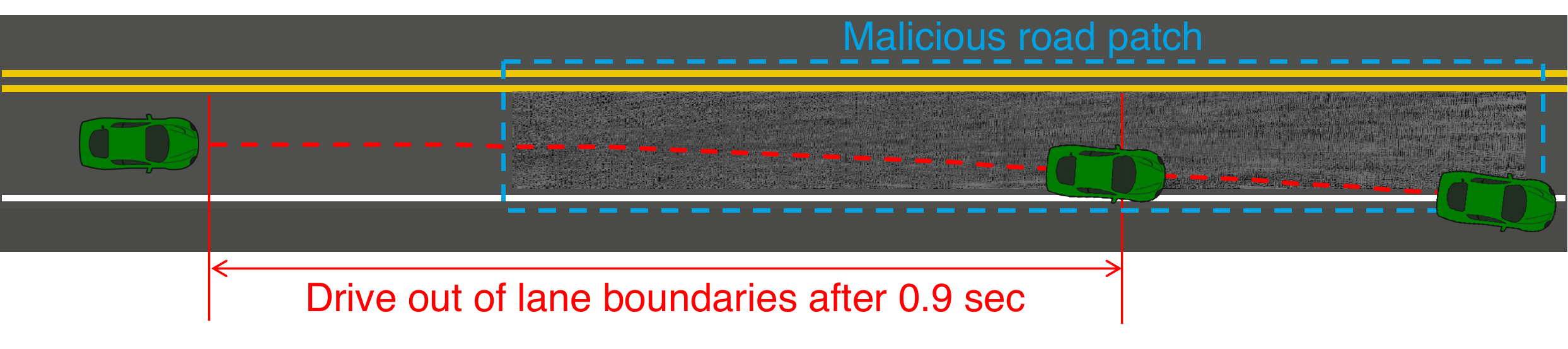}
\caption{Malicious road patch and car trajectory for comma2k19-1.}
\label{fig:patch}
\end{figure}

We evaluate our method on a state-of-the-art open-source LKAS, OpenPilot~\cite{openpilot}, which is reported to have similar performance as Tesla Autopilot and GM Super Cruise, and better than all other manufacturers.~\cite{openpilotreview}.
It adopts a general DNN-based lane detection pipeline as shown in Fig.~\ref{fig:openpilot_LKAS}. Particularly, it implements a DNN model with recursive layers and uses the standard Model Predictive Control (MPC)~\cite{MPC} as the lateral controller.
We evaluate our method on 3 scenarios and the results are summarized in Table~\ref{tbl:attack_result}. The comma2k19-1 and comma2k19-2 are real-world highway scenarios selected from the comma2k19 dataset~\cite{comma2k19}. The LGSVL-1 is a simulated highway scenario created by LGSVL, an industry-grade photo-realistic Autonomous Driving simulator. As shown, our attack succeeds to cause the victim vehicle to drive out of the highway lane boundaries (over 0.745 meters deviations) within 1.3 seconds, which is much smaller than the average driver reaction time (2.3 seconds)~\cite{driver_reaction_time2000}. Fig.~\ref{fig:patch} shows an example malicious road patch generated by our method.

\begin{figure}[!t]
\centering
\includegraphics[width=3.4in]{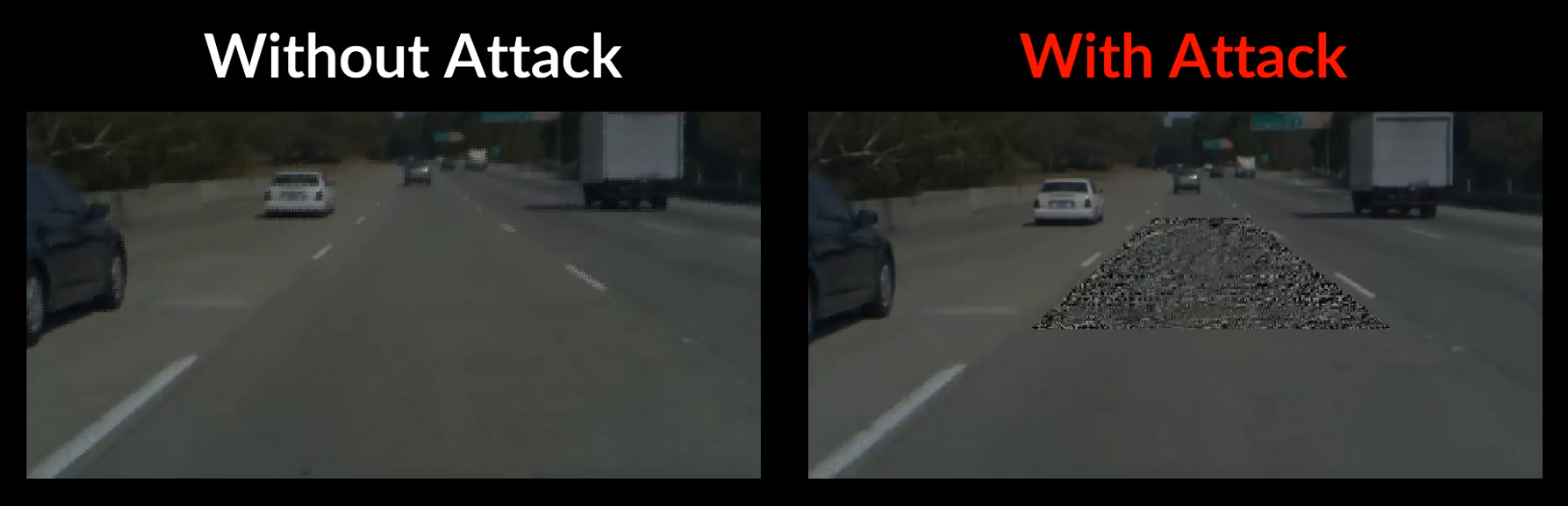}
\caption{A snapshot of our attack demo from the victim driver's view at our project website \textbf{\url{https://sites.google.com/view/lane-keeping-adv-attack/}}. In this demo, our attack is able to cause the victim car to drive off a highway lane after only 0.9 seconds.}
\label{fig:demo}
\end{figure}

To better illustrate the attack effect, we prepare a short demo from the victim car driver's view at our project website \textbf{\url{https://sites.google.com/view/lane-keeping-adv-attack/}}. In this demo, our attack causes the victim car to drive off a highway lane after only 0.9 seconds. Fig.~\ref{fig:demo} shows a snapshot of the demo. The demo video is synthesized from the transformed camera image via the car motion model base input generation, i.e., by placing our malicious road patch on the BEV image, generating the camera inputs from the BEV, and updating the next frame state based on the car motion model.

\begin{table}[tbp]
\caption{Attack effectiveness when the deviation goal is 0.745 meters (driving off lane boundaries on the highway).}
\label{tbl:attack_result}
\centering
\begin{tabular}{@{}cccc@{}}
\toprule
Scenario    & Avg. Speed        & Attack Time     & Patch Size (W $\times$ L) \\ \midrule
comma2k19-1 & 126 km/h (78 mph) & 0.9 s           & 3.6 m $\times$ 36 m         \\
comma2k19-2 & 105 km/h (65 mph) & 1.0 s           & 3.6 m $\times$ 36 m         \\
LGSVL-1     & 72 km/h (45 mph)  & 1.3 s           & 3.6 m $\times$ 36 m         \\ \bottomrule
\end{tabular}
\end{table}

\section{Conclusion and Future Plans}
In this work, we design and implement the first systematic approach to attack real-world DNN-based LKASes. We evaluate our approach on a state-of-the-art LKAS and our preliminary results show that our attack can successfully cause it to drive off lane boundaries within as short as 1.3 seconds. In the future, we plan to (1) perform more comprehensive evaluation including more diverse scenarios, different car types, other DNN lane detection models, (2) demonstrate the attack in real-world experiments, and (3) design effective defenses.



%
\bibliographystyle{IEEEtran}
\bibliography{main.bib}

\end{document}